\documentclass[twocolumn,eqsecnum,floats,aps,nofootinbib,showpacs,preprintnumbers]{revtex4}

\usepackage{amsmath,amssymb,amsfonts}
\usepackage{graphicx}

\usepackage{hyperref} 

\def\be{\begin{equation}}
\def\ee{\end{equation}}
\def\ba{\begin{eqnarray}}
\def\ea{\end{eqnarray}}
\def\nn{\nonumber}

\def\g{\gamma}
\def\d{\delta}

\def\k{\kappa}

\def\m{\mu}

\def\p{\pi}

\def\r{\rho}
\def\vr{\varrho}

\def\ph{\phi}

\def\D{\Delta}

\newcommand{\Pl}{\ell_{\rm Pl}} 
\newcommand{\muzero}{\m^o} 
\newcommand{\mubar}{{\bar \m}} 
\newcommand{\K}{{\cal K}}

\newcommand{\secref}[1]{Section~\ref{#1}}

\newcommand{\eqnref}[1]{(\ref{#1})}
\newcommand{\figref}[1]{FIG.~\ref{#1}}
\newcommand{\appref}[1]{Appendix~\ref{#1}}

\begin{document}

\preprint{IGPG-07/2-5}

\title{Effective Dynamics for the Cosmological Bounces\\
in Bianchi Type I Loop Quantum Cosmology}
\author{Dah-Wei Chiou}
\affiliation{
Institute for Gravitational Physics and Geometry,
Physics Department,
The Pennsylvania State University, University Park, PA 16802, U.S.A.}

\begin{abstract}
The detailed formulation for loop quantum cosmology (LQC) in the Bianchi I model with a massless scalar field was recently constructed. In this paper, its effective dynamics with the LQC discreteness corrections is studied and the equations of motion are analytically solved, showing that the big bang singularity is replaced by the big bounces, which take place up to three times, once in each diagonal direction, whenever each of the area scale factors approaches its critical value in the Planck regime measured by the reference of the scalar field momentum.
\end{abstract}

\pacs{04.60.Kz, 04.60.Pp, 98.80.Qc, 03.65.Sq}

\maketitle


\section{Introduction}
The comprehensive formulation for loop quantum cosmology (LQC) in the spatially flat-isotropic model has been constructed \cite{Ashtekar:2006uz,Ashtekar:2006rx}. With a massless scalar field serving as the \emph{emergent time}, the result shows that the quantum evolution is \emph{deterministic across the deep Planck regime} and in the backward evolution of the states which are semiclassical at late times, \emph{the big bang is replaced by a big bounce}. Based on the same principles, the construction was further improved by a more direct implementation of the underlying physical ideas of loop quantum gravity (LQG) \cite{Ashtekar:2006wn}. In the improved dynamics, \emph{the big bounce occurs precisely when the matter density enters the Planck regime}, regardless of the value of the momentum $p_\ph$ of the scalar field.

Both the precursor strategy (``$\m_o$-scheme'') and the improved strategy (``$\mubar$-scheme'') were applied and reconstructed for the Bianchi I model to include anisotropy \cite{Chiou:2006qq}. The analytical investigation shows that the state in the kinematical Hilbert space associated with the classical singularity is \emph{completely decoupled} in the difference evolution equation, indicating that the classical singularity is resolved in the quantum evolution and the big bounce may take place when any of the area scales undergoes the vanishing behavior.

While a thorough numerical investigation remains to be done to draw the definite conclusion for the details of the quantum evolution in the Bianchi I model, this paper studies its effective dynamics with LQC discreteness corrections. Not only does the result affirm the anticipations in \cite{Chiou:2006qq} but more intuitive pictures are also obtained in this semiclassical approach, giving an insight into how and why the big bounces take place.

In accordance with the formulation in \cite{Chiou:2006qq}, this paper focus specifically on the model with a massless scalar field. In the context of effective dynamics with LQC discreteness corrections, the similar analysis for more generic Bianchi I models with the inclusion of arbitrary matter with equation of state $w<+1$ is also investigated in \cite{Chiou:2007sp}, which gives the similar results for the occurrence of big bounces with only difference in detail. With arbitrary matter sources, however, the equations of motion are very complicated and a proper approximation has to be used. By contrast, in the special case of a massless scalar field, the equations of motion can be solved analytically and therefore the underlying physics is more transparent.

This paper is organized as follows. In \secref{sec:classical dynamics}, the classical dynamics of the Bianchi I cosmology with a massless scalar source is solved in terms of Ashtekar variables in Hamiltonian formulation. The effective dynamics with LQC corrections in $\mubar$-scheme is constructed and solved in \secref{sec:mubar dynamics}. Its phenomenological ramifications are discussed in \secref{sec:discussion}. As a comparison to the $\mubar$-scheme, the effective dynamics in $\m_o$-scheme is also included in \appref{sec:muzero dynamics}.

\section{Classical Dynamics}\label{sec:classical dynamics}
The spacetime metric of Bianchi type I is given as:
\be
ds^2=-dt^2+a_1^2(t)dx^2+a_2^2(t)dy^2+a_3^2(t)dz^2.
\ee
In terms of Ashtekar variables, the phase space of Bianchi I models is given by the diagonal triad
variables $p_I$ and diagonal connection variables $c_I$ for
$I=1,2,3$, which satisfy the canonical relation:
\be \{c_I,p_J\}=8\p G\g\,\d_{IJ}.
\ee
The triad variables $p_I$ are
related with the length scale factors $a_I$ via:
\be\label{eqn:p and a}
p_1=a_2a_3,\qquad
p_2=a_1a_3,\qquad p_3=a_1a_2.
\ee
In the presence of a massless scalar field $\ph(\vec{x},t)=\ph(t)$,
(which is independent of the spatial coordinates with homogeneity assumed),
the classical dynamics is govern
by the Hamiltonian constraint:
\ba\label{eqn:cl Hamiltonian}
&&C=C_{\rm grav}+C_\ph\\
&=&-\frac{\big(c_2p_2c_3p_3+c_1p_1c_3p_3+c_1p_1c_2p_2\big)}{8\p G\g^2\sqrt{{p_1p_2p_3}}}
+\frac{p_\ph^2}{2\sqrt{p_1p_2p_3}},\nn
\ea
where $p_\ph$ is the conjugate momentum of $\ph$ and has the canonical relation with $\ph$:
\be
\{\ph,p_\ph\}=1.
\ee

We can simplify the Hamiltonian by choosing the lapse function $N=\sqrt{p_1p_2p_3}$
and thus introducing the new time variable $dt'=(p_1p_2p_3)^{-1/2}dt$.
The rescaled Hamiltonian constraint is given by
\be\label{eqn:cl rescaled Hamiltonian}
H=-\frac{\left(c_2p_2c_3p_3+c_1p_1c_3p_3+c_1p_1c_2p_2\right)}{8\p G\g^2}
+\frac{p_\ph^2}{2}.\\
\ee

The equations of motion are governed by the Hamilton's equations:
\ba
\label{eqn:cl eom 1}
\frac{dp_\ph}{dt'}&=&\{p_\ph,H\}=0\quad\Rightarrow\quad
p_\ph\ \text{is constant}\\
\label{eqn:cl eom 2}
\frac{d\ph}{dt'}&=&\{\ph,H\}=p_\ph,\\
\label{eqn:cl eom 3}
\frac{dc_1}{dt'}&=&\{c_1,H\}=8\p G\g\,\frac{\partial\, H}{\partial p_1}\nn\\
&=&-\g^{-1}c_1\left(c_2p_2+c_3p_3\right),\\
\label{eqn:cl eom 4}
\frac{dp_1}{dt'}&=&\{p_1,H\}=-8\p G\g\,\frac{\partial\, H}{\partial c_1}\nn\\
&=&\g^{-1}p_1\left(c_2p_2+c_3p_3\right),
\ea
and so on for $c_2$, $c_3$, $p_2$, $p_3$ in the cyclic manner.
In addition to the Hamilton's equations, the constraint that the Hamiltonian must vanish yields
\ba\label{eqn:cl eom 5}
&&H(c_I,p_I)=0\quad
\Rightarrow\\
p_\ph^2&=&\frac{1}{4\p G\g^2}
\big(c_2p_2c_3p_3+c_1p_1c_3p_3+c_1p_1c_2p_2\big).\nn
\ea

Combining \eqnref{eqn:cl eom 3} and \eqnref{eqn:cl eom 4} gives
\ba\label{eqn:const Ki}
\frac{d}{dt'}(p_Ic_I)=0,\quad\Rightarrow\quad
p_Ic_I=8\p G\g\hbar\,\K_I,
\ea
where $\K_I$ are dimensionless constants, which will be used to
parameterize the solutions of evolution.
Taking \eqnref{eqn:const Ki} into \eqnref{eqn:cl eom 5}, we have
\be\label{eqn:p_ph and K}
p_\ph^2=16\p G\hbar^2
\left\{\K_2\K_3+\K_1\K_3+\K_1\K_2\right\}
\ee
or equivalently
\be\label{eqn:K}
\K_\ph^2=2
\left(\K_2\K_3+\K_1\K_3+\K_1\K_2\right),
\ee
if we define
\be\label{eqn:def of p_ph}
p_\ph:=\hbar\sqrt{8\p G}\,\K_\ph.
\ee

Putting \eqnref{eqn:const Ki} into \eqnref{eqn:cl eom 4} gives
\be
\frac{1}{p_1}\frac{dp_1}{dt'}={8\p G \hbar}\left(\K_2+\K_3\right),
\ee
By referring to \eqnref{eqn:cl eom 2}, this leads to
\be\label{eqn:cl diff eq 2}
\frac{1}{p_I}\frac{dp_I}{d\ph}=8\p G\hbar\,\frac{\K_2+\K_3}{p_\ph}
=\sqrt{8\p G}\,\Big(\frac{1-\k_I}{\k_\ph}\Big),
\ee
where we scale the parameters $\K_I=\K\k_I$, $\K_\ph=\K\k_\ph$ such that
\be\label{eqn:para constraint}
\k_1+\k_2+\k_3=1,
\qquad
\k_1^2+\k_2^2+\k_3^2+\k_\ph^2=1.
\ee
Regarding $\ph$ as the \emph{emergent time}, the solutions of evolution are given by
\be\label{eqn:cl sol 1}
p_I(\ph)=p_I(\ph_0)\,e^{\sqrt{8\p G}\big(\frac{1-\k_I}{\k_\ph}\big)(\ph-\ph_0)},
\ee
or equivalently
\be\label{eqn:cl sol 2}
a_I(\ph)=a_I(\ph_0)\,e^{\sqrt{8\p G}\,\frac{\k_I}{\k_\ph}(\ph-\ph_0)}.
\ee

The classical Bianchi I model with a massless scalar field admits both ``Kasner-like''
(two of $\k_I$ positive and the other negative) and ``Kasner-unlike''
(all $\k_I$ positive) solutions.
The Kasner-like solution, which has two expanding and one contracting directions (say $\k_\ph>0$),
eventually encounters the ``Kasner-like singularity''
(a given regular cubical cell stretches as an infinitely long line) in the far past
and the ``planar collapse'' (a regular cubical cell stretches as an infinitely large plane)
in the far future. On the other hand, the Kasner-unlike solution, with all directions
expanding, encounters the ``Kasner-unlike singularity''
(a regular cubical cell vanishes to a point) in the far past and no
planar collapse.

We will see that with LQC discreteness corrections, both Kasner-like and Kasner-unlike singularities are resolved and replaced by the \emph{big bounces},
whereas the planar collapse remains its destiny even one of the three diagonal directions
approaches infinitely small length scale.

\section{Effective Dynamics in $\mubar$-Scheme}\label{sec:mubar dynamics}
In LQC, the connection variables $c_I$ do not exist and should be replace by
holonomies. In the effective theory, to capture the quantum corrections, following the procedures used in the isotropic case \cite{Taveras:IGPG preprint,Singh:2005xg}, we take the
prescription to replace $c_I$ by $\sin(\mubar_Ic_I)/\mubar_I$, introducing discreteness
variables $\mubar_I$. In the improved strategy ($\mubar$-scheme) used in
Bianchi I LQC \cite{Chiou:2006qq}, $\mubar_I$ are not fixed constants but given by
\be
\mubar_1=\sqrt{\frac{\D}{p_1}}\,,\quad
\mubar_2=\sqrt{\frac{\D}{p_2}}\,,\quad
\mubar_3=\sqrt{\frac{\D}{p_3}}\,,
\ee
where $\D=\frac{\sqrt{3}}{2}(4\p\g\Pl^2)$ is the \emph{area gap} in the full theory of LQG.

Imposing this prescription plus the loop quantum correction to the inverse triad on \eqnref{eqn:cl Hamiltonian}, we have the effective Hamiltonian constraint to the leading order:
\ba\label{eqn:qm Hamiltonian original}
C_{\rm eff}&=&f(p_1)f(p_2)f(p_3)\frac{p_\ph^2}{2}
-\frac{f(p_1)f(p_2)f(p_3)}{8\p G \g^2}\\
&&\quad\times
\left\{
\frac{\sin(\mubar_2c_2)\sin(\mubar_3c_3)}{\mubar_2\mubar_3}p_2p_3+
\text{cyclic terms}
\right\},\nn
\ea
where $f(p_I)$ is the eigenvalue of the inverse triad operator $\widehat{1/\sqrt{p_I}}$. The loop quantization gives the quantum corrections:
\be
f(p_I)\sim
\left\{
\begin{array}{cr}
\frac{1}{\sqrt{{p_I}}}\left(1+{\cal O}(\Pl^2/p_I)\right) & \text{for}\ p_I\gg\Pl^2\\
\propto{p_I}^n/\Pl^{2n+1} & \text{for}\ p_I\ll\Pl^2
\end{array}
\right.
\ee
with the Planck length $\Pl:=\sqrt{G\hbar}$ and a positive $n$. The corrections to $f(p_I)$ is significant only in the Planckian region in the vicinity of $p_I=0$. From now on, we will ignore the quantum corrections to $f(p_I)$ by simply taking its classical function $f(p_I)=p_I^{-1/2}$. [We will see that in the backward evolution the big bounce takes place much earlier before the discreteness correction on the inverse triad operator becomes considerable, and it is the ``non-locality'' effect (i.e., using the holonomies) that accounts for the occurrence of the big bounce.]

With $f(p_I)=p_I^{-1/2}$, by choosing $dt'=(p_1p_2p_3)^{-1/2}dt$, the Hamiltonian constraint \eqnref{eqn:qm Hamiltonian original} can be rescaled as
\ba\label{eqn:qm Hamiltonian}
&&H_\mubar=\frac{p_\ph^2}{2}\\
&&\quad
-\frac{1}{8\p G \g^2}
\left\{
\frac{\sin(\mubar_2c_2)\sin(\mubar_3c_3)}{\mubar_2\mubar_3}p_2p_3+
\text{cyclic terms}
\right\}.\nn
\ea
Again, the equations of motion are given by the Hamilton's equations and
the constraint that the Hamiltonian must vanish:
\ba
\label{eqn:qm eom 1}
\frac{dp_\ph}{dt'}&=&\{p_\ph,H_\mubar\}=0\quad\Rightarrow\quad
p_\ph\ \text{is constant}\\
\label{eqn:qm eom 2}
\frac{d\ph}{dt'}&=&\{\ph,H_\mubar\}=p_\ph,\\
\label{eqn:qm eom 3}
\frac{dc_1}{dt'}&=&\{c_1,H_\mubar\}=8\p G\g\,\frac{\partial\, H_\mubar}{\partial p_1}\nn\\
&=&-\g^{-1}
\left(\frac{3\sin(\mubar_1c_1)}{2\mubar_1}-\frac{c_1\cos(\mubar_1c_1)}{2}\right)\nn\\
&&\quad\ \times
\left(\frac{\sin(\mubar_2c_2)}{\mubar_2}p_2
+\frac{\sin(\mubar_3c_3)}{\mubar_3}p_3\right),\\
\label{eqn:qm eom 4}
\frac{dp_1}{dt'}&=&\{p_1,H_\mubar\}=-8\p G\g\,\frac{\partial\, H_\mubar}{\partial c_1}\nn\\
&=&\g^{-1}p_1\cos(\mubar_1c_1)\nn\\
&&\quad\ \times
\left(\frac{\sin(\mubar_2c_2)}{\mubar_2}p_2
+\frac{\sin(\mubar_3c_3)}{\mubar_3}p_3\right),
\ea
and
\ba\label{eqn:qm eom 5}
&&H_\mubar(c_I,p_I)=0\quad
\Rightarrow\qquad p_\ph^2=\\
&&\frac{1}{4\p G\g^2}
\left\{
\frac{\sin(\mubar_2c_2)\sin(\mubar_3c_3)}{\mubar_2\mubar_3}p_2p_3+
\text{cyclic terms}
\right\}.\nn
\ea
[Note that in the classical limit $\mubar_Ic_I\rightarrow0$, we have
$\sin(\mubar_Ic_I)/\mubar_I\rightarrow c_I$,
$\cos(\mubar_Ic_I)\rightarrow1$ and therefore
\eqnref{eqn:qm eom 3}--\eqnref{eqn:qm eom 5} reduce to their
classical counterparts \eqnref{eqn:cl eom 3}--\eqnref{eqn:cl eom 5}.]

By \eqnref{eqn:qm eom 3} and \eqnref{eqn:qm eom 4}, we have
\ba\label{eqn:qm dpc/dt'}
&&\left(\frac{3\sin(\mubar_Ic_I)}{2\mubar_I}-\frac{c_I\cos(\mubar_Ic_I)}{2}\right)\frac{dp_I}{dt'}
+p_I\cos(\mubar_Ic_I)\frac{dc_I}{dt'}\nn\\
&&=\frac{d}{dt'}\left[p_I\frac{\sin(\mubar_Ic_I)}{\mubar_I}\right]=0,
\ea
which gives
\be\label{eqn:qm pc}
p_I\frac{\sin(\mubar_Ic_I)}{\mubar_I}
=8\p G\g\hbar\,\K_I.
\ee\\
Taking \eqnref{eqn:qm pc} into \eqnref{eqn:qm eom 5} again
gives the same constraints on the constant parameters as in
\eqnref{eqn:p_ph and K} or \eqnref{eqn:K}.

Substituting \eqnref{eqn:qm pc} into \eqnref{eqn:qm eom 4} yields
\be\label{eqn:qm diff eq 1}
\frac{1}{p_1}\frac{dp_1}{dt'}
=8\p G \hbar\,\cos(\mubar_1c_1)(\K_2+\K_3).
\ee
By regarding $\ph$ as the emergent time via \eqnref{eqn:qm eom 2} and expressing $\cos x=\pm\sqrt{1-\sin^2 x}$,
\eqnref{eqn:qm diff eq 1} then gives
\be\label{eqn:qm diff eq 2}
\frac{1}{p_I}\frac{dp_I}{d\ph}=
\pm\sqrt{8\p G}\,\Big(\frac{1-\k_I}{\k_\ph}\Big)
\left[1-\frac{\vr_I}{\vr_{I\!,\,{\rm crit}}}\right]^{1/2},
\ee
where we define the \emph{directional density}:
\be
\varrho_I:=\frac{p_\ph^2}{p_I^3}
\ee
for the $I$-direction and its critical value is given by the \emph{Planckian matter density}
$\r_{\rm Pl}$ times a numerical factor:
\be
\vr_{I\!,\,{\rm crit}}:=\left(\frac{\k_\ph}{\k_I}\right)^2\r_{\rm Pl},
\qquad
\r_{\rm Pl}:=(8\p G \g^2\D)^{-1}.
\ee

\begin{widetext}

\begin{figure}
\begin{picture}(400,150)(0,0)

\put(-60,0){
\begin{picture}(460,150)(0,0)


\put(10,133){(a)}
\put(175,133){(b)}
\put(345,133){(c)}


\resizebox{\textwidth}{!}
{\includegraphics{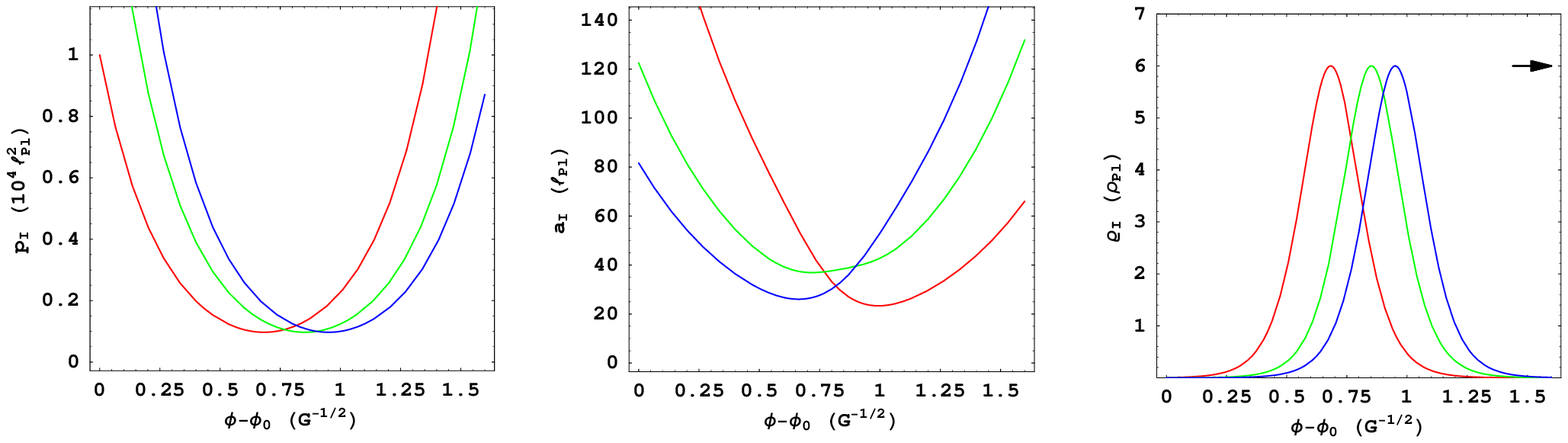}}

\end{picture}
}

\end{picture}
\caption{$\k_1=\k_2=\k_3=1/3$, $\k_\ph=\sqrt{2/3}$; $p_1(\ph_0)=1\times10^4\Pl^2$,
$p_2(\ph_0)=2\times10^4\Pl^2$, $p_3(\ph_0)=3\times10^4\Pl^2$;
and $p_\ph=2\times10^3\hbar\sqrt{8\p G}$ (i.e., $\K\k_\ph=2\times10^3$).
The
{red} lines are for $p_1$, $a_1$, $\varrho_1$;
{green} for $p_2$, $a_2$, $\varrho_2$;
and
{blue} for $p_3$, $a_3$, $\varrho_3$. The values of
$\varrho_{1,\,{\rm crit}}$, $\varrho_{2,\,{\rm crit}}$
and $\varrho_{3,\,{\rm crit}}$ are pointed by the arrow(s) in (c). (The Barbero-Immirzi parameter is set to $\g=1$.)}\label{fig:fig1}
\end{figure}

\begin{figure}
\begin{picture}(400,150)(0,0)

\put(-60,0){
\begin{picture}(460,150)(0,0)


\put(10,133){(a)}
\put(175,133){(b)}
\put(345,133){(c)}


\resizebox{\textwidth}{!}
{\includegraphics{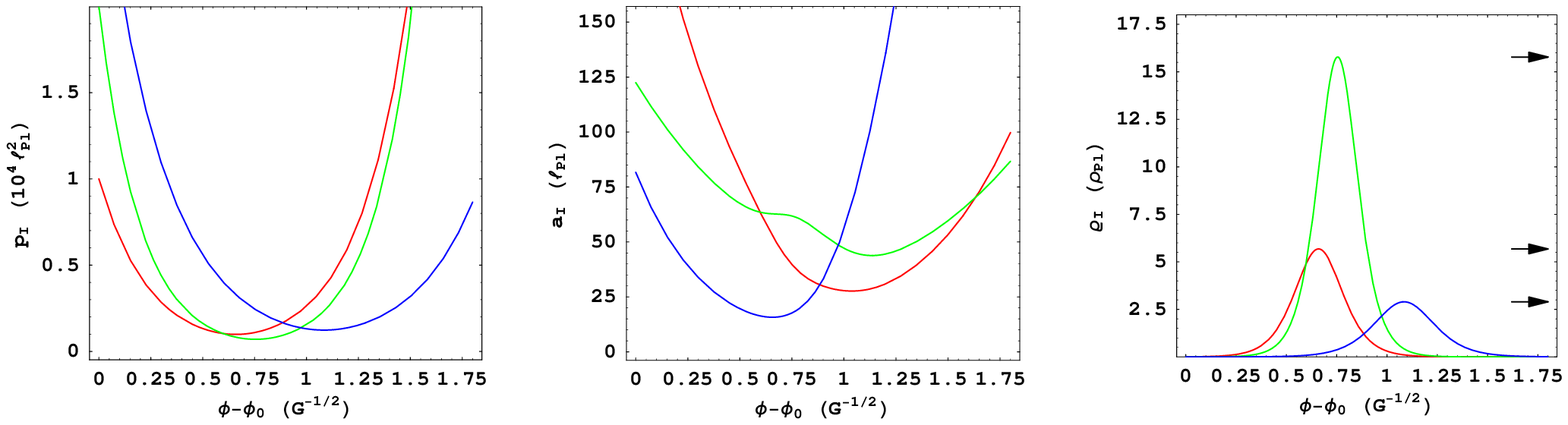}}

\end{picture}
}

\end{picture}
\caption{$\k_1=1/3$, $\k_2=1/5$, $\k_3=7/15$, $\k_\ph=\sqrt{142}/15$; $p_1(\ph_0)=1\times10^4\Pl^2$,
$p_2(\ph_0)=2\times10^4\Pl^2$, $p_3(\ph_0)=3\times10^4\Pl^2$;
and $p_\ph=2\times10^3\hbar\sqrt{8\p G}$ (i.e., $\K\k_\ph=2\times10^3$).}\label{fig:fig2}
\end{figure}

\begin{figure}
\begin{picture}(400,150)(0,0)

\put(-60,0){
\begin{picture}(460,150)(0,0)


\put(10,133){(a)}
\put(175,133){(b)}
\put(345,133){(c)}


\resizebox{\textwidth}{!}
{\includegraphics{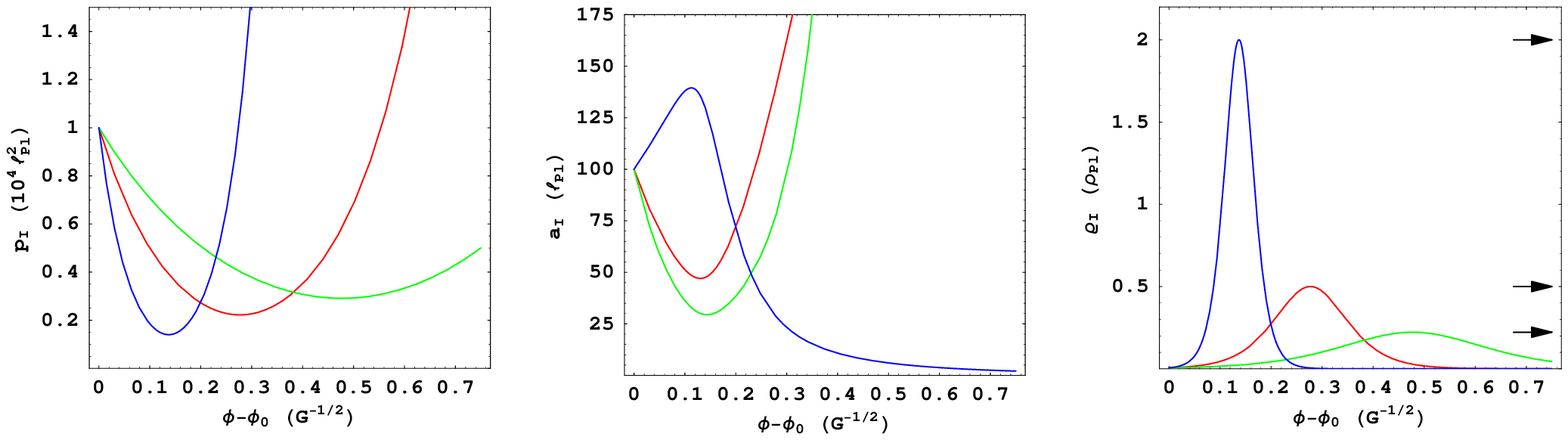}}

\end{picture}
}

\end{picture}
\caption{$\k_1=1/2$, $\k_2=3/4$, $\k_3=-1/4$, $\k_\ph=1/\sqrt{8}$; $p_1(\ph_0)=p_2(\ph_0)=p_3(\ph_0)
=1\times10^4\Pl^2$;
and $p_\ph=2\times10^3\hbar\sqrt{8\p G}$ (i.e., $\K\k_\ph=2\times10^3$).}\label{fig:fig3}
\end{figure}

\begin{figure}
\begin{picture}(400,150)(0,0)

\put(-60,0){
\begin{picture}(460,150)(0,0)


\put(10,133){(a)}
\put(175,133){(b)}
\put(345,133){(c)}


\resizebox{\textwidth}{!}
{\includegraphics{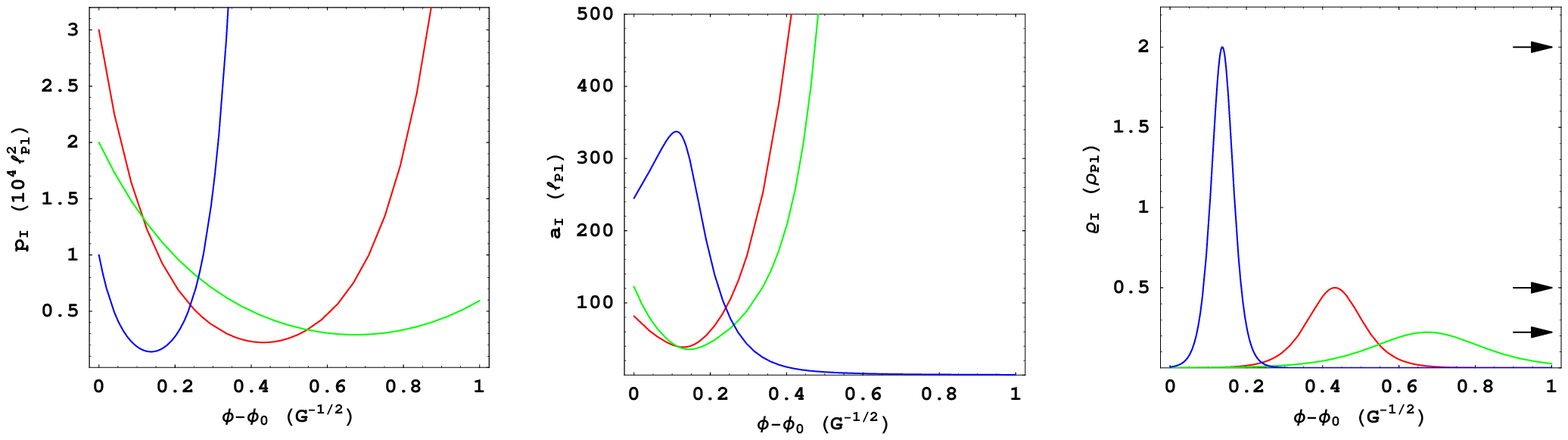}}

\end{picture}
}

\end{picture}
\caption{$\k_1=1/2$, $\k_2=3/4$, $\k_3=-1/4$, $\k_\ph=1/\sqrt{8}$; $p_1(\ph_0)=3\times10^4\Pl^2$,
$p_2(\ph_0)=2\times10^4\Pl^2$, $p_3(\ph_0)=1\times10^4\Pl^2$;
and $p_\ph=2\times10^3\hbar\sqrt{8\p G}$ (i.e., $\K\k_\ph=2\times10^3$).}\label{fig:fig4}
\end{figure}

\end{widetext}

\section{Discussion}\label{sec:discussion}
As opposed to the classical equation \eqnref{eqn:cl diff eq 2}, in which $p_I$ continues to decrease toward the classical singularity in the backward evolution, the effective equation in \eqnref{eqn:qm diff eq 2} flips sign exactly at the moment when $\vr_I$ approaches its critical value $\vr_{I\!,\,{\rm crit}}$.
Note that by \eqnref{eqn:p and a} $p_I$ can be regarded as the \emph{area} scale factors.
Therefore, with the LQC discreteness corrections, \eqnref{eqn:qm diff eq 2} shows that  the singularities (both Kasner-like and Kasner-unlike) are resolved and replaced by the big bounces in the backward evolution when any of the area scales undergoes the vanishing behavior. Across the bounces, the equation of motion again comes closer and closer to the classical counterpart. Hence, the semiclassicality is retained on both asymptotic sides of the evolution.

Furthermore, the detailed evolutions of $p_I$ are decoupled in different diagonal directions and evolve independently of one another once the initial conditions ($p_I(\ph_o)$, $p_\ph$ and $\k_I$) are specified. Thus, the bounces occur up to three times, once in each direction, whenever each of the directional densities $\vr_I$ approaches its critical value.
As expected, in $\mubar$-scheme, the critical values $\vr_{I\!,\,{\rm crit}}$ are in the Planck regime of ${\cal O}(\hbar\Pl^{-4})$ and \emph{independent of the value of $p_\ph$} ($\vr_{I\!,\,{\rm crit}}$ depend on $p_\ph$ only through the ratio $\k_\ph/\k_I\equiv\K_\ph/\K_I$).\footnote{In \appref{sec:muzero dynamics}, the old precursor strategy ($\m_o$-scheme) is presented and it shows that the critical value of $\vr_I$ can be made arbitrarily small by increasing $p_\ph$.}
Note that $\vr_I$ have the same dimension as the matter density $\r:=p_\ph^2/(2p_1p_2p_3)$ and $\vr_I$ play the same role as $\r$ does in the isotropic case, signaling the occurrence of big bounces.

On the other hand, the planar collapse is \emph{not} resolved but one of the length scale factors $a_I$ continues the vanishing behavior in the Kasner-like case. This is
expected since the classical solutions \eqnref{eqn:const Ki} and \eqnref{eqn:cl sol 1} yield $\mubar_Ic_I\rightarrow0$ (and $\muzero_Ic_I\rightarrow0$ in $\m_o$-scheme) toward the planar collapse and therefore the quantum corrections become more and more negligible (in both schemes).

For given initial conditions, the differential equation \eqnref{eqn:qm diff eq 2} can be solved numerically. The behaviors of $p_I(\ph)$, $a_I(\ph)$ and $\vr_I(\ph)$ are depicted in parts (a), (b) and (c) respectively in \figref{fig:fig1} and \figref{fig:fig2} for Kasner-unlike solutions and in \figref{fig:fig3} and \figref{fig:fig4} for Kasner-like solutions.

The fact that smallness of $p_I$ (not of $a_I$) is an indication of the occurrence of big bounces seems to support the suggestion that ``area is more fundamental than length in LQG'', although whether this is simply a technical artifact or reflects some deep physics is still not clear. (See Section VII.B of \cite{Rovelli:1997yv} for some comments on this aspect and \cite{Rovelli:1993vu} for more details.)
Meanwhile, as the length operator has been shown to have a
discrete spectrum \cite{Thiemann:1996at}, the fact that the
vanishing of the length scale factor in the planar collapse is not stopped seems to contradict the discreteness of the length spectrum. Whether we miss some important ingredients when imposing the fundamental discreteness of LQG in the LQC construction or indeed area is more essential than length remains an open question for further investigation.

It is also noteworthy that \eqnref{eqn:qm diff eq 2} remains invariant if we rescale $p_\ph\rightarrow l^3 p_\ph$ and $p_I\rightarrow l^2p_I$ at the same time. This is reminiscent of the idea as suggested in \cite{Rovelli:1990ph,Rovelli:1992vv} that area is measurable only if the surface is \emph{coupled with the material reference}. The scaling invariance, however, breaks down in the full LQC theory since the quantum evolution is governed by a difference equation \cite{Chiou:2006qq}, in which the step size of difference introduces an additional scale in the deep Planck regime.\footnote{Therefore, the semiclassicality is retained in the full quantum theory only for large $p_\ph$ and $p_I$. Accordingly, we put big values of $p_\ph$ and $p_I(\ph_o)$ in the figures to make sense of the semiclassical approach for the effective dynamics. The figures are trivially rescaled under the scaling.}

Meanwhile, related to the above observation, the physical meaning of the directional densities $\vr_I$ can be interpreted as the (inverse of) area scales, again, \emph{measured by the reference of the matter content}. The big bounces take place whenever one of the area scales becomes very small by the reference of the matter momentum. It is then attempting to regard not only $\ph$ as the ``internal clock'' (emergent time) but also $p_\ph$ as the ``internal rod'' --- namely, the measurement of both temporal and spatial geometries makes sense only in the presence of matter content. This observation may support the ideas of the relational interpretation of quantum mechanics with real rods and clocks such as studied in \cite{Gambini:2006ph} (see also \cite{Rovelli:1990ph,Rovelli:1992vv}), although the link is far from clear. If this concept is taken seriously, in return, we might be able to further improve the $\mubar$-scheme to better reflect the underlying physics of LQG such that the difference equation of evolution in the full LQC theory also respects the scaling invariance mentioned above.

\begin{acknowledgements}
The author would like to thank Abhay Ashtekar, Golam Hossain, Tomasz Pawlowski, Parampreet Singh for useful discussions and especially Kevin Vandersloot for sharing his private notes and important ideas. This work was supported in part by the NSF grant PHY-0456913.
\end{acknowledgements}


\appendix

\section{Effective Dynamics in $\m_o$-Scheme}\label{sec:muzero dynamics}
One of the virtues of the improved strategy ($\mubar$-scheme) in the isotropic model is to fix the serious drawback in the old precursor strategy ($\m_o$-scheme) that the critical value of the matter density at which the bounce occurs can be made arbitrarily small by increasing the momentum $p_\ph$.

As the directional densities $\vr_I$ play the same role in the Bianchi I model as the matter density $\r$ does in the isotropic case, we expect that the critical value of $\vr_I$ at which the bounce occurs can be made arbitrarily small by increasing the momentum $p_\ph$ in $\m_0$-scheme but is independent of $p_\ph$ in $\mubar$-scheme. The latter is what is shown in the main text of this paper. For comparison, the effective dynamics in $\m_o$-scheme is presented here.

In the effective theory of $\m_o$-scheme, we take the
prescription to replace $c_I$ by $\sin(\muzero_Ic_I)/\muzero_I$ with the \emph{fixed} numbers $\muzero_I$ for discreteness. Analogous to \eqnref{eqn:qm Hamiltonian},
we have the effective (rescaled) Hamiltonian constraint:
\ba
&&H_{\m_o}=\frac{p_\ph^2}{2}\\
&&\quad
-\frac{1}{8\p G \g^2}
\left\{
\frac{\sin(\muzero_2c_2)\sin(\muzero_3c_3)}{\muzero_2\muzero_3}p_2p_3+
\text{cyclic terms}
\right\}.\nn
\ea

Again, the equations of motion are given by the Hamilton's equations and
the constraint that the Hamiltonian must vanish:
\ba
\label{eqn:qm0 eom 1}
\frac{dp_\ph}{dt'}&=&\{p_\ph,H_{\m_o}\}=0\quad\Rightarrow\quad
p_\ph\ \text{is constant}\\
\label{eqn:qm0 eom 2}
\frac{d\ph}{dt'}&=&\{\ph,H_{\m_o}\}=p_\ph,\\
\label{eqn:qm0 eom 3}
\frac{dc_1}{dt'}&=&\{c_1,H_{\m_o}\}=8\p G\g\,\frac{\partial\, H_{\m_o}}{\partial p_1}\nn\\
&=&-\g^{-1}
\left(\frac{\sin(\muzero_1c_1)}{\muzero_1}\right)\nn\\
&&\quad\ \times
\left(\frac{\sin(\muzero_2c_2)}{\muzero_2}p_2
+\frac{\sin(\muzero_3c_3)}{\muzero_3}p_3\right),\\
\label{eqn:qm0 eom 4}
\frac{dp_1}{dt'}&=&\{p_1,H_{\m_o}\}=-8\p G\g\,\frac{\partial\, H_{\m_o}}{\partial c_1}\nn\\
&=&\g^{-1}p_1\cos(\muzero_1c_1)\nn\\
&&\quad\ \times
\left(\frac{\sin(\muzero_2c_2)}{\muzero_2}p_2
+\frac{\sin(\muzero_3c_3)}{\muzero_3}p_3\right),
\ea
and
\ba\label{eqn:qm0 eom 5}
&&H_{\m_o}(c_I,p_I)=0\quad
\Rightarrow\qquad p_\ph^2=\\
&&\frac{1}{4\p G\g^2}
\left\{
\frac{\sin(\muzero_2c_2)\sin(\muzero_3c_3)}{\muzero_2\muzero_3}p_2p_3+
\text{cyclic terms}
\right\}.\nn
\ea

From \eqnref{eqn:qm0 eom 3} and \eqnref{eqn:qm0 eom 4}, we have
\be\label{eqn:qm0 dpc/dt'}
\frac{d}{dt'}\left[p_I\frac{\sin(\muzero_Ic_I)}{\muzero_I}\right]=0,
\ee
which gives
\be\label{eqn:qm0 pc}
p_I\frac{\sin(\muzero_Ic_I)}{\muzero_I}
=8\p G\g\hbar\,\K_I.
\ee
Taking \eqnref{eqn:qm0 pc} into \eqnref{eqn:qm0 eom 5}
gives the same constraints on the constant parameters as in
\eqnref{eqn:p_ph and K} or \eqnref{eqn:K}.

Substituting \eqnref{eqn:qm0 pc} into \eqnref{eqn:qm0 eom 4} yields
\be\label{eqn:qm0 diff eq 1}
\frac{1}{p_1}\frac{dp_1}{dt'}
=8\p G \hbar\,\cos(\muzero_1c_1)(\K_2+\K_3).
\ee
By \eqnref{eqn:qm0 eom 2} and $\cos x=\pm\sqrt{1-\sin^2x}$,
\eqnref{eqn:qm0 diff eq 1} leads to
\be\label{eqn:qm0 diff eq 2}
\frac{1}{p_I}\frac{dp_I}{d\ph}=
\pm\sqrt{8\p G}\,\Big(\frac{1-\k_I}{\k_\ph}\Big)
\left[1-
\left(\frac{\varrho_I}{\varrho^{\m_o}_{I\!,\,{\rm crit}}}\right)^{2/3}\right]^{1/2},
\ee
which gives the bouncing solutions with the behaviors similar to those given by \eqnref{eqn:qm diff eq 2} except that the critical value of $\vr_I$ at which the big bounce takes place is given by
\be
\varrho^{\m_o}_{I\!,\,{\rm crit}}:=
\left[
\left(\frac{\k_\ph}{\k_I}\right)^2\frac{\r_{\rm Pl}\D}{{\muzero_I}^2}
\right]^{3/2}
\frac{1}{p_\ph},
\ee
which can be made arbitrarily small by increasing the value of $p_\ph$. This tells that $\m_o$-scheme gives wrong semiclassical behavior and should be improved by $\mubar$-scheme.

\end{document}